# Wind Tunnel Testing and Modeling Implications of an Advanced Turbine Cascade

Dr. Sharath Sathish

Thermal Systems Laboratory, Department of Mechanical Engineering

Indian Institute of Science, Bangalore 560 012, India

**Abstract**

This paper describes the extensive Wind Tunnel (WT) linear cascade testing campaign carried out on a constant section turbine blade developed for low subsonic applications. Comprehensive experimental program was designed to determine the aerodynamic behaviour of this blade under a wide range of varying geometrical like Pitch / Chord Ratio, Stagger Angle and Aerodynamic like Reynolds-Roughness, Mach number, Incidence angle conditions. In addition to the classical Two-Dimensional (2D) measurements, targeted Three-Dimensional (3D) surveys have been performed to complement the 2D results, allowing to draw supplementary conclusions as to the behaviour of the considered blade in three-dimension. The performed experiments predominantly covered the transitional and beginning of a fully turbulent flow regime (maximum Re = 2.5 million). Post processing of experimental results were done in view of exploitation and judgment of some key aerodynamic aspects - parameters such as profile section load distribution, loss coefficient and flow deviation angle. Alongside with confirmation of the design targets, a parallel goal pursued in the current WT testing, was to get a more accurate idea as to the predictive capabilities of the employed CFD tools (CFX and MISES) under the basic flow condition generated in the WT. The aim was to get a global picture, as to the level of agreement that can be reached by the used design tools while accounting for the afore-mentioned design, operational variables. The obtained measurement results clearly indicated the success of the design in achieving the pre-set targets. The paper further provides a detailed discussion on the identified discrepancies: measurement versus predictions, of the key parameters at the geometric – aerodynamic design as well as off-design conditions.

| Nomenclature | | Subscripts & Superscripts | |
|---|---|---|---|
| alpha | exit flow angle (°) | 01 | total quantity at inlet plane |
| inc. | incidence angle (°) | 02 | total quantity at outlet plane |
| $I_{tu}$ | turbulence intensity (%) | 2 | static quantity at outlet plane |
| ks | sandwall roughness ($micron$) | Greek symbols | |
| P | pressure ($bar$) | α | deviation angle (°) |
| pc | pitch-to-chord ratio | θ | stagger angle (°) |
| PS | pressure side | | |
| Mach | Mach number | | |
| Re | Reynold's number | | |
| SMD | surface Mach distribution | | |
| SMW | smooth wall | | |
| SS | suction side | | |
| WT | wind tunnel | | |
| s | specific entropy ($J/kg.K$) | | |
| $Y_2$ | total pressure coefficient $(P_{01} - P_{02})/(P_{02} - P_2)$ | | |

## 1. Introduction

From the early days the turbine blading development activities have been intimately linked with WT measurements [1–5]. Experimental determination of the exit flow angle, profile loss coefficient, traditionally expressed as drag coefficient were among the major targets of such measurements. Despite incredible advances in the CFD tools as well as the availability of highly flexible and fast 2D blade sections and 3D aerofoil generators, WT testing continue to be a very useful element complementing the design process.

Indeed, as compared to highly complex flow structure under real turbine operation, WT testing provides an opportunity for generating the simplest type of Two-Dimensional and Three-Dimensional stationary cascade flows, the knowledge and understanding of which is a pre-requisite for a successful handling of aerodynamic phenomenon within real machines. At the same time capabilities for accurate prediction of such type of simplified flows constitute an essential element in demonstrating the much-needed initial confidence in the existing aerodynamic design tools and hence are of vital importance to the major manufacturers. Such predictive abilities are not only instrumental in validating the design tools but also constitute a valuable supplement to better understand the flow physics, interpret the results of experimental test rigs and support their respective performance modelling.

The cornerstone of this paper is to elucidate the degree of agreement of CFD and measurement which have been achieved while considering the impact of Reynolds number, surface roughness and transitional phenomenon, compressibility, the incoming flow incidence, pitch / chord ratio and stagger angle on the key parameters such as section loading distribution, profile loss and flow deviation angle. The organization of the paper elucidates the linear cascade test setup, discussion of the test results, the numerical modelling and the comparison of test and numerical results.

**2. Development of a new profile**

A new high efficiency profile section has been developed for axial turbine applications, based on Reaction Technology. The intent of this development activities was to generate experimental evidence as to the behavior of the selected profile section loading philosophy for design applications in the high pressure, e.g., low aspect ratio section of the industrial turbines which contribute a significant portion of the overall turbine power output. The targeted blades operate predominantly under low subsonic flow conditions with the applied overarching aerodynamic design principles based on [6]. Extensive iterations of the blade design were carried out with MISES [7], Quasi-3D flow solver followed by 3D CFD using Ansys CFX over an identified design space range to arrive at the optimum design. The design optimization has been performed to attain a compromise profile featuring not only the highest possible efficiency under the design condition but also give rise to a very robust behavior throughout the entire design space. During the design process, the optimization methods based on FORMA tool [8], and the Class and Shape Transformation based optimization technique [9], were explored.

## 3. Description of the Wind-Tunnel and linear cascade test setup

The experimental campaign was performed in the Transonic Wind Tunnel (WT) for Linear Cascades (Figure 3.1) located at the Laboratory of Fluid Machines (LFM) of Politecnico di Milano (Italy). The facility is a blow down type supplied by pressurized air storage of approximately 5000 Kg of dehydrated air (dew point -13°C) at 180 bar. The maximum flow rate available at the facility is approximately 8 kg/s. The test section main dimensions are 450 mm in pitch wise direction and 80 mm in blade height. The cascades used for the present research were all composed by a sufficiently high number of blades (8 blades); a blade passage was always instrumented by 35 pressure taps located on the mid span section, 17 on the blade pressure side (PS) and 18 on the blade suction side (SS). Main characteristics of blades and cascade are reported in Table 3.1.

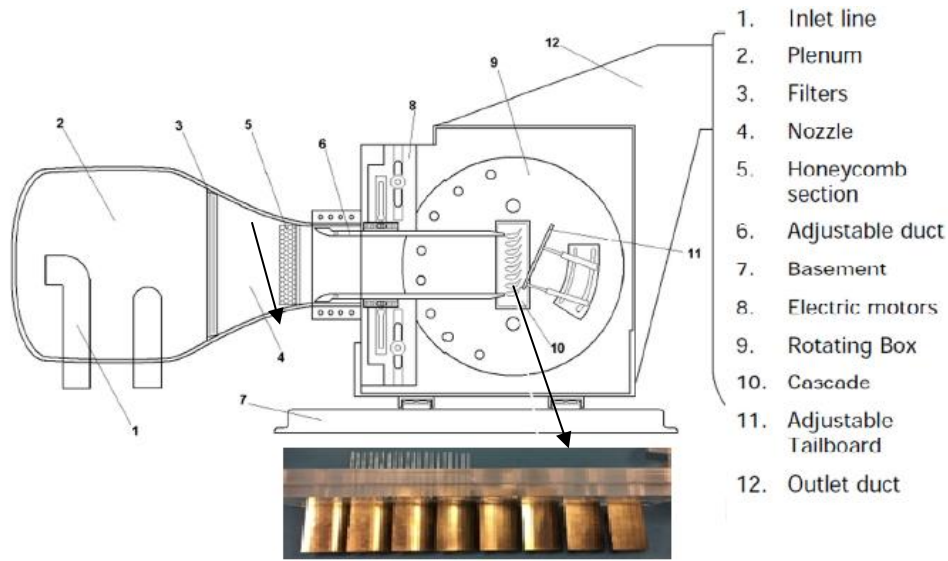

Figure 3.1 Schematic of the wind tunnel linear-cascade facility

| Parameter | Unit | Value |
|---|---|---|
| No. of blades | - | 8 |
| Blade height | mm | 80.0 |
| Blade chord | mm | 60.0 |
| Axial chord | mm | 41.9 |
| Blade pitch | mm | 49.3 |
| Stagger angle | degree | 44.9 |
| Inlet flow angle | degree | 6.5 |
| Surface roughness | micron | 0.3-0.4 |
| Test-section pitch wise dimension | mm | 450.0 |

Table 3.1 Blade and nominal cascade characteristics

### 3.1. Instrumentation and calibration

Two of the eight blades have been instrumented with 35 pressure taps on the mid span section, 17 located on one blade pressure side (PS) and 18 on the suction side (SS) of a different blade. The blade taps exhibit a diameter of 0.4 mm and are connected to the pressure transducers by a hole running inside of the blade (i.e., parallel to the blade span) mounting on the blade base a stainless steel glued to the pressure line. The instrumented blades have been designed to pass through the holding plate so to easily connect the stainless-steel tube coming from the pressure taps to the pressure transducers. To properly locate the instrumented passage in correspondence of the measuring section, the cascade has been designed to locate three adjacent passing-through blades in the center section of the cascade. The other five blades have been installed on the holding blade by means of calibrated pivots. The holding plate have been instrumented also by 20 pressure taps at the endwall to cover three central blade pitches for the aim of monitoring the isentropic Mach number at the beginning of the tests. The downstream flow periodicity is ensured by a movable tailboard located downstream of the cascade and verifying through preliminary 5-hole probe pitch wise traversing. The facility is supported by an automatic traversing system for different kinds of directional probes, both upstream and downstream the cascade. A wedge type 3-hole probe is installed upstream of the blades to characterize the incoming flow field at midspan. A miniaturized directional 5-hole probe was used for pitchwise and spanwise traversal at 50% of the axial chord, downstream of the cascade. The 5-hole probe provides the local 3D downstream flow field. Each measuring traverse included a minimum of 50 measuring points along the two interested pitches. Several traverses at the cascade inlet were performed to verify and guarantee the pitchwise uniformity of the incoming the flow field. To support 3D measurements performed on a full secondary plane downstream of the cascade, a detailed survey of the upstream boundary layer was performed by means of a flattened total pressure probe one axial chord upstream of the cascade leasing edges for every different geometrical and Mach number test condition.

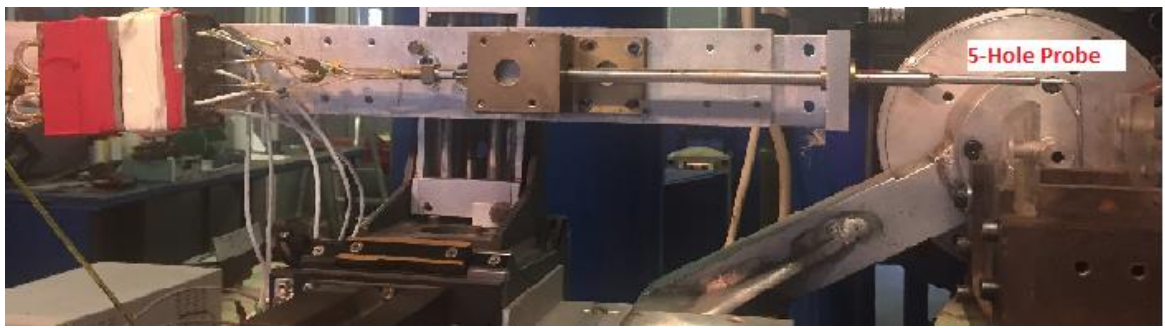

Figure 3.2 5-hole probe and calibration setup (Credits: LFM, Polimi)

The key instrumentation used for aerodynamic measurements in the wind tunnel are described here. The 5-hole probe (Figure 3.2) is applied to traverse the flow field downstream of the cascade and allows for the measurement of the local 3D field viz., local total pressure, static pressure (i.e., local Mach number) and flow angles (yaw and pitch). The probe can be placed at a given distance from the cascade trailing edge position and traversed in both axial and spanwise direction (for the aim of 3D tests, i.e., secondary flow survey). The probe is equipped with 5 pre-calibrated pressure transducers one for each of the sensor lines. All transducers are operated in fully differential mode so to reduce measurement uncertainties. The sensing line coming from the central hole is measured as differential with respect to the upstream total pressure. Since the probe is pre-oriented with the expected outlet flow angle, the differential pressure provides the local total pressure drop throughout the cascade. Nevertheless, the application of the calibration matrix allows for the determination of the local total pressure. Pressure taps have been manufactured to exhibit a hole diameter of 0.2 mm. The sensing line at the exit is perpendicular to the cone curtain air.

The 5-hole probe is calibrated for the following range:

- Yaw angle: ± 24°
- Pitch Angle: ±16°
- Mach number : 0.2-1.0

The calibration procedure allows the definition of one calibration surface for every considered calibration Mach number. On every calibration surface four coefficients are defined for the aim of deriving the four local flow field unknowns: *Pt, P, Yaw, Pitch*.

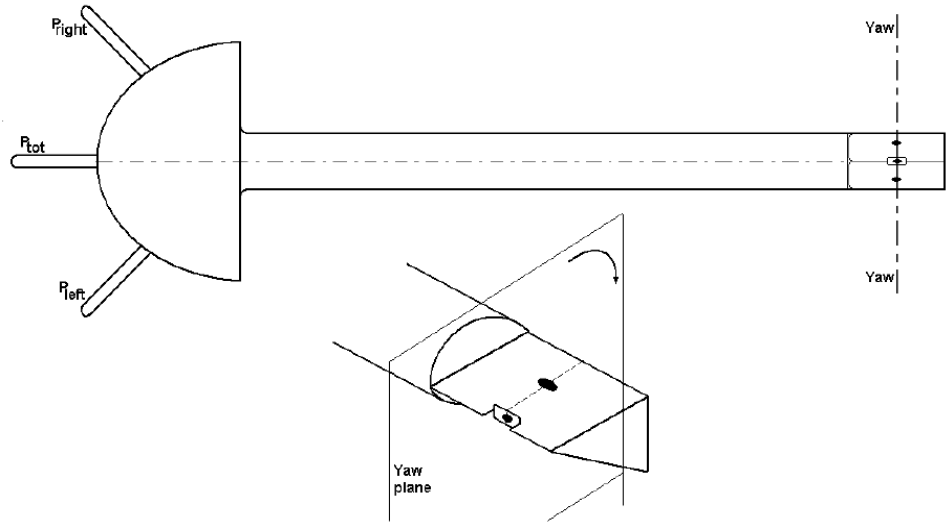

Figure 3.3 Schematic of 3-hole probe (Credits: LFM, Polimi)

The 3-hole probe used for the present campaign is a wedge edge shaped probe like the one reported in Figure 3.3. This kind of probe is suitable for the survey of 2D flow fields. The probe has been calibrated in an angular yaw range of ± 24° and for Mach number range up to 0.9. The 3-hole probe is used to provide the reference flow field at midspan during the tests. By the application of the calibration surfaces, the 3-hole probe allows to measure the local values of total pressure, static pressure, and flow angle. During testing, the 3-hole probe is pre-aligned with the direction of the inlet flow at the entrance of the cascade. The reading of the central hole of the cascade is a direct measurement of the local total pressure of the flow. The pressure line connected to the probe central tap is then split in two lines: one goes directly to a multi-transducer box while the second one is used as reference pressure for the differential reading of the central hole of the 5-hole probe. The lateral taps of the 3-hole probe are directly connected to the multi transducer box. The multi-transducer box is a rake of 30 Kulite transducers, characterized by a full scale of 50 PSI. The pressure transducers provide several pressure readings related to the three readings of the upstream 3-hole probe and to the 20 readings of the pressure taps located the end wall of the cascade holding plate. Every transducer has been calibrated in the relevant application range. The pressure scanner is pneumatic scanner connecting 48 pressure sensors to a single pressure transducer by means of an electrically actuated rotating disk. It is used to scan the 35 pressure taps located at the midspan.

*3.2. Upstream flow characterization*

The natural flow field upstream of the cascade has been characterized by total pressure probe and hot-wire probe traversing, both in tangential and radial direction. Since the natural Tu intensity at the cascade inlet is lower than 1%, for the aim of the present test, Tu generator capable of producing a minimum Tu intensity of 5% has been preliminarily designed and verified by Tu measurements. The Tu generator is composed of a grid of cylindrical bars. The main features of the Tu generator are listed below and its placement in the cascade is shown in Figure 3.4.

- Bars diameter: 5.5 mm.
- Bar spacing: 16 mm.
- Grid position: 250 mm upstream of the cascade leading edges .
- Bars orientation: spanwise.
- Grid orientation: tangential.

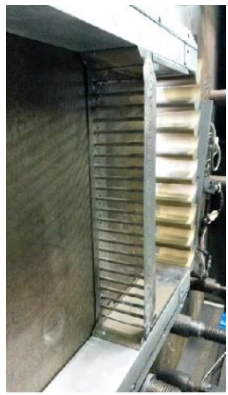

Figure 3.4 Turbulence generator in the wind tunnel (Credits: LFM, Polimi)

The flow resulting from the application of the turbulent generator has been characterized in terms of time averaged flow field and streamwise turbulence component on a plane located 60 mm (one blade chord) upstream of the cascade leading edge. This corresponds to a distance of approximately 35 bar diameters downstream of the Tu generator. This datum guarantees that the mixing out process and most relevant part of the Tu decay downstream of the rake of bars has already taken place in the measuring section. Thus, the measured conditions can be assumed extremely close to the ones experienced by the cascade, in terms of both turbulent and averaged flow field.

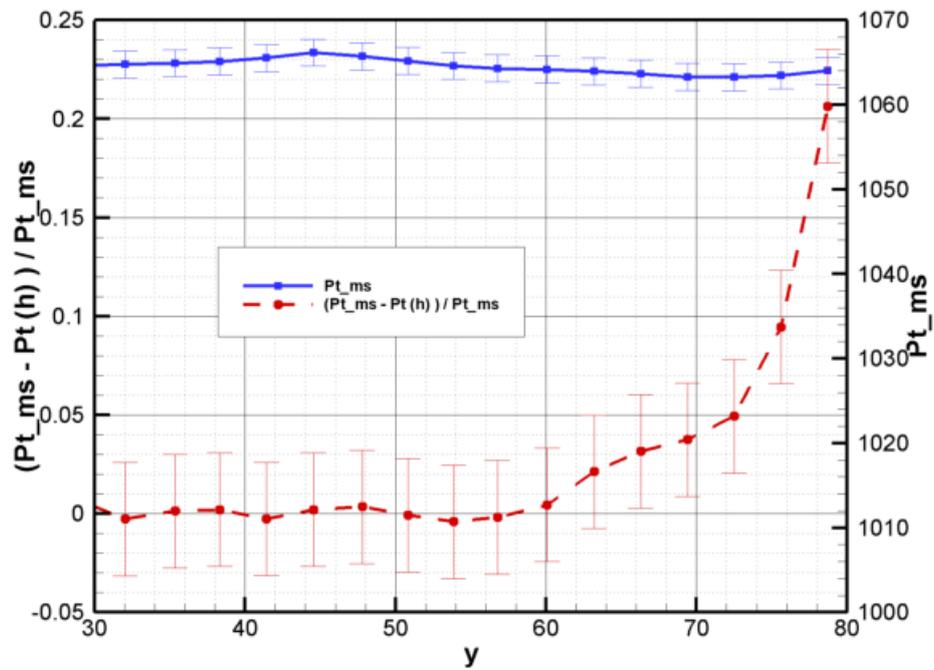

Figure 3.5 Total pressure tangential distribution upstream of the cascade as measured by pitot tube

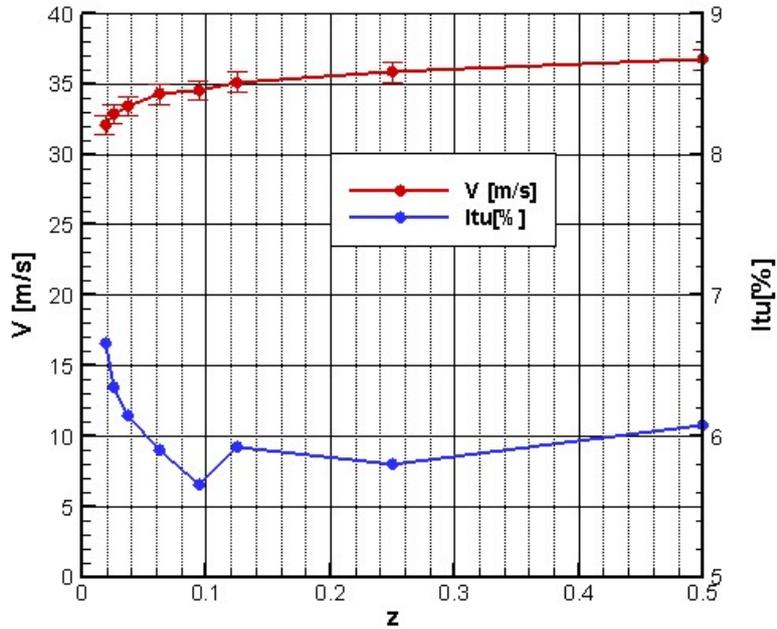

Figure 3.6 Velocity and Turbulence intensity spanwise distribution upstream of the cascade as measured by hot-wire anemometer

Flow survey has been performed with an isentropic cascade outlet Mach number of 0.3. The upstream flow characterization results are detailed below.

- <u>Total pressure distribution (from pitot tube):</u> The quantities that are measured are i) the time averaged total pressure distribution along the tangential direction ($Pt\_ms$; blu), ii) non-dimensional total pressure drop ($\Delta Pt/Pt\_ms$, red) along the tangential direction. Total pressure measurements, obtained by traversing a flattened total pressure probe, show a good tangential uniformity and a fully developed upstream boundary layer (Figure 3.5).
- <u>Velocity and Turbulence measurements (from hot-wire):</u> The quantities that are measured are i) spanwise distribution of the velocity along blade midspan (red), ii) spanwise distribution of turbulence intensity($Itu\%$, blu). Both hot-wire measurements have been performed by the application of a single wire probe, allowing the measurement of the axial turbulent component $v'$ and time averaged axial velocity $V$ at 60 mm upstream of the cascade. The radial distribution shows a flat region extending from z=0.5 to approximately z=0.1characterized by flat uniform value of *Itu*=6%. In the last 10% of the blade height, *Itud* increases rapidly, reaching a value of 15% (Figure 3.6).

## 3.3. Cascade test procedure

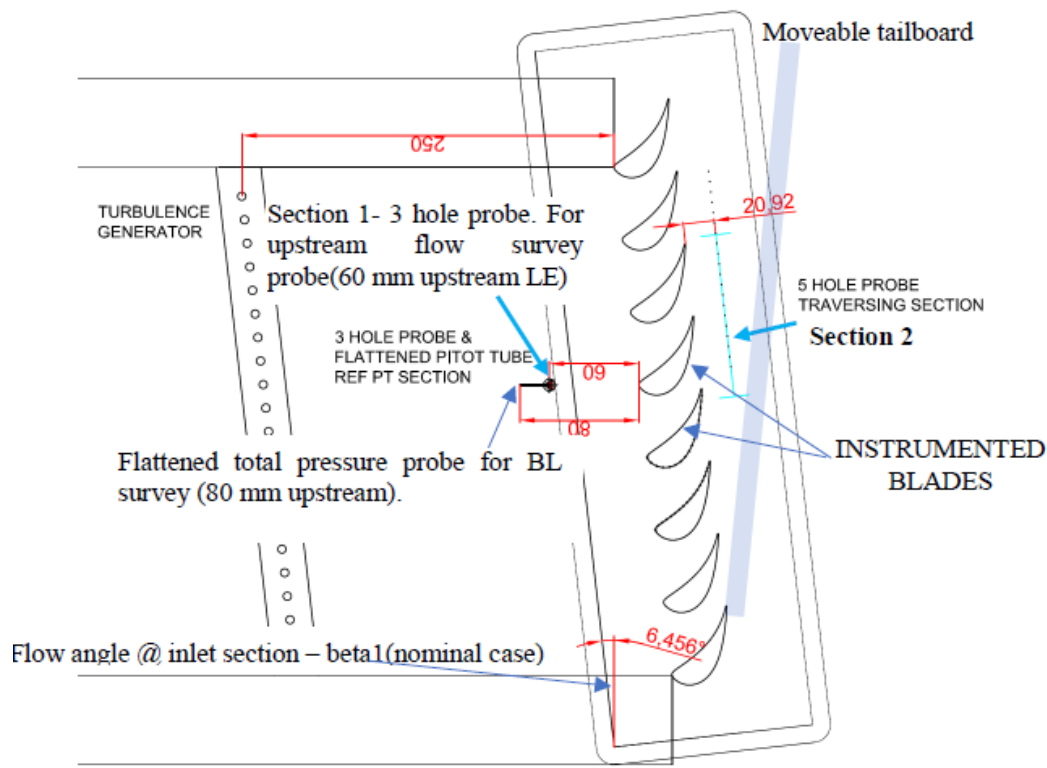

Figure 3.7 Sketch of the cascade nominal incidence configuration

The 2D and 3D flow field survey is performed at downstream, upstream and profile surface locations. At the downstream section (Section-2 in Figure 3.7), 5-hole probe is located at 0.5 axial chord from blade trailing edge. The midspan measurement generates the 2D flow field measurements, while the traversal of the probe from midspan to top endwall generates 3D flow field measurements. The measured quantities are total pressure ($P_{t2}$), yaw angle from tangential direction ($\beta_2$) and static pressure ($P_{s2}$). At the upstream section (Section-1 in Figure 3.7), 3-hole probe is located at one axial chord from the blade leading edge at midspan. The measured quantities are total pressure ($P_{t1}$), yaw angle from tangential direction ($\beta_1$) and static pressure ($P_{s1}$). Profile pressure distribution on the blades is measured by 18 pressure taps on the suction side and 17 pressure taps on the pressure side, all located at blade midspan. The assessment of a satisfactory flow periodicity downstream of trailing edge, for the desired outlet Mach number, is obtained by the angular setting of the downstream tailboard. A preliminary set of tests is performed before every single test at a given outlet Mach number, to set the proper tailboard angle. Flow periodicity is then verified by tangentially traversing the 5-hole probe along two pitches and checking the periodicity between the two pitches in terms of the following distributions.

- Pitchwise outlet flow angle distribution ($\beta_2$)
- Pitchwise outlet Mach number distribution ($M_2$)
- Pitchwise distribution of the local total pressure loss coefficient ($Y_{loc2}$)
- Pitchwise static pressure distribution ($Ps_2$)

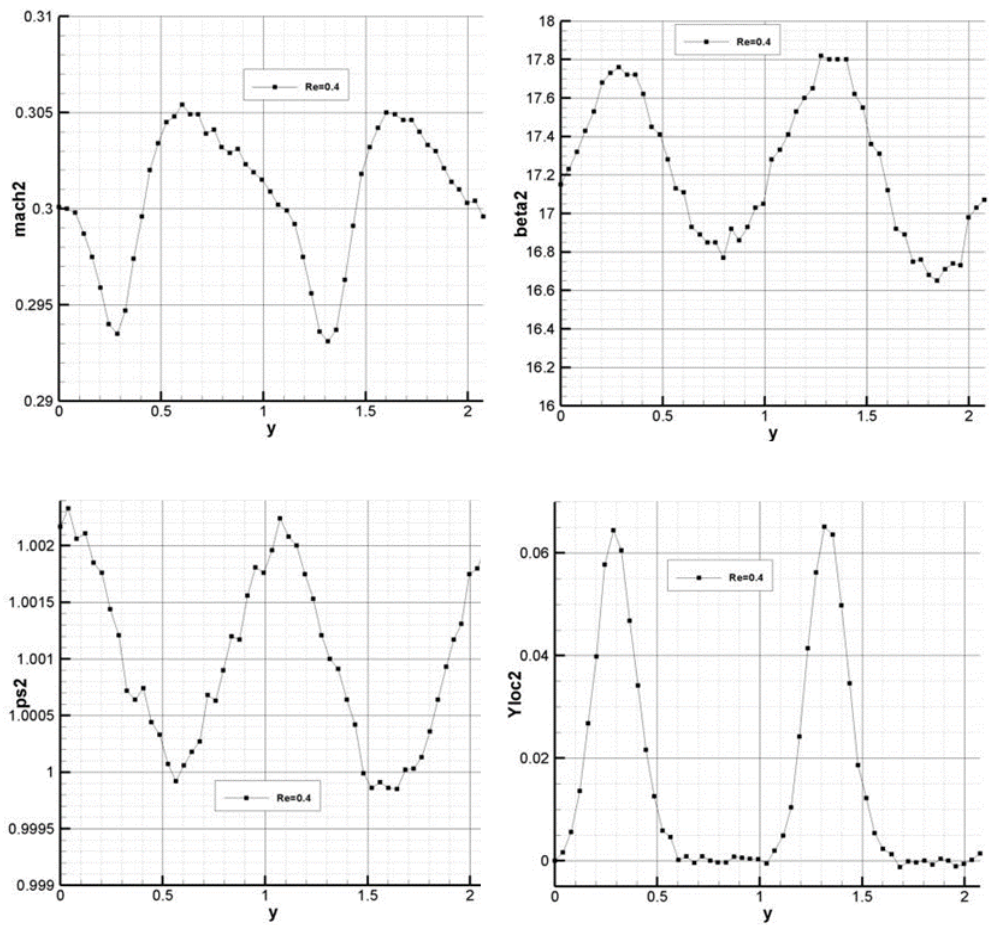

Figure 3.8 Achieved periodicity of flow parameters in the cascade test

The flow distribution plots (Figure 3.8) plots establishes the excellent periodicity achieved between the two adjacent blade pitches. The central portion of the reported traverse refers to the passage delimited by the instrumented blades. The plots evidence only very slight differences between the two pitches. Wakes, clearly represented in the Mach number and loss distribution, evidence identical distribution: the peaks of loss and the Mach defects exhibits identical values. The maximum difference in static pressure is contained in less than 50 Pa, i.e., below the measurement uncertainty. The same occurs for the outlet angle, whose maximum difference is below 0.2°. The $Re_2$ variation is obtained by the pressurization of the test section by creating choked flow conditions at the wind tunnel discharge section. The sonic throat dimension can be fine-tuned by two control valves, thus allowing the independent establishment of both predefined $Re_2$ and $M_2$. The maximum operating pressure was 3.5 bar at the cascade outlet section. Moreover, 20 pressure taps at the endwall, covering the three central blade passages, were also used to initially set flow periodicity and $M_2$ and $Re_2$ numbers. Data provided during tests by the two pressure probes have been mass averaged along two blade pitches to provide 2D and 3D blade performance, mainly in terms of outlet angle and loss coefficients. To reduce the loss coefficient uncertainty, the 5-hole probe was always pre-aligned with the downstream averaged flow angle. The central hole reading was directly connected to a differential high accuracy pressure transducer which senses the prevailing upstream total pressure reading coming from the 3-hole probe. To further reduce uncertainty, 5 to 15 repetitions (depending on $Re_2$ and $M_2$) of the same test have been performed. As a result, the total pressure coefficient ($Y$) uncertainty was evaluated with a confidence level of 95%.

*3.4. Cascade test objectives*

The intent of cascade test activities was to generate experimental evidence as to the behavior of the optimized profile section loading philosophy for design applications in the high pressure, low aspect ratio section of the industrial turbines. The high pressure stages contribute a significant portion of the overall turbine power output. The targeted blades operate predominantly under low subsonic flow conditions with the applied overarching aerodynamic design principles enunciated in the previous chapters. Extensive iterations of the blade design were carried out with MISES quasi-3D blade-to-blade flow solver followed by 3D CFD using Ansys CFX over an identified design space range to arrive at the optimum design. The design optimization has been performed to attain a profile featuring not only the highest possible efficiency under the design condition but also give rise to a very robust behavior throughout the entire design space. The wide-ranging tests involved variation of exit Reynolds number from 0.4 million to 2.5 million, at three exit Mach numbers of 0.3, 0.5 and 0.7, the Reynolds and Mach numbers being independently controlled. A full survey of the cascade sensitivity to off-design incidence in the range ±40° was performed for exit Mach number equal to 0.3, including detailed 3D tests at the nominal incidence as well as at incidences equal to ±40°. For the same exit Mach number of 0.3, additional 2D and 3D tests were also carried out for two extreme values of solidity (±20% of the design value) and one different stagger angle (7° more tangential with respect to the design one).

## 4. CFD Methodology

With the double aim of assessing the prediction capability of the tools in use by the Company for blade aerodynamic design, and of integrating the information coming from the test campaign, computational models of the flow in the cascade were reproduced for most of the tested configurations by resorting to a Computational Fluid Dynamic (CFD) model. The numerical model is based on the commercial finite-volume solver ANSYS-CFX. Steady flow is assumed systematically in all the computed conditions; quasi-3D simulations, considering a straight streamtube around midspan, were performed for all the conditions of interest; fully 3D simulations (i.e., including half of the blade channel from midspan to the endwall) were also carried out for some conditions of specific interest. In all the cases, the equations are discretized using the available high-resolution TVD scheme for the inviscid fluxes and central differences for the viscous terms. The equations are solved over structured multi-block meshes composed by hexahedral elements. Even though a look-up-table approach based on the IAPWS-IF97 is routinely applied in the design phase, tests were performed in air in ambient (or nearly ambient) condition, so the fluid is treated as polytropic ideal gas.

Turbulence effects were introduced by resorting to several models; at first, the basic fully turbulent k-ω model was selected, considering both hydraulically smooth and rough wall [10]; then, to include the effects of transition observed in the experiments, the correlation-based model $\gamma$-$Re_\theta$ was considered in combination to the k-ω [11]. By virtue of those choices, special care was made to the construction of the near-wall mesh, so to ensure wall $y^+$ below unity for all the conditions of interest (even in the less refined meshes used for the grid dependence analysis). Boundary conditions were assigned as total quantities, flow direction, turbulence intensity and eddy viscosity ratio at the inlet, while static pressure was imposed at the cascade exit. All the above values were available from the experiments (except the eddy viscosity ratio, set equal to 1, which is reasonable value for wind-tunnel flows) and were assigned accordingly, so to match the exact Mach and Reynolds number conditions tested.

At the beginning of the computational campaign a dedicated grid dependence study was carried out to define the mesh resolution required to achieve a grid independent solution. To this end, quasi-3D simulations were performed with progressively increasing cell number starting from 50 k cells, all of them featuring wall $y^+$ below unity and considering fully turbulent flow; the study indicated that meshes composed by 250 k cells are required to obtain a grid independent solution. In case of fully 3D simulations, grids composed by 5 million of cells were used, featuring wall $y^+$ below unity also on the endwall surface.

## 5. Linear cascade test results and discussions

*5.1. Analysis of section loading distribution at nominal incidence*

The blade loading distribution at nominal and off-nominal conditions are presented followed by the loading distributions for varying *p/c* (inverse of solidity) and stagger angle. The section loading, as characterized by surface pressure or Mach number distributions, are compared, and shown here. The way such distributions compare with predictions, in particular the diffusion rate in the unguided zone and whether the flow has a tendency of separation, are of primary importance. Figure 5.1 compares the surface Mach distributions at 1.0 million $Re_2$ for 0.3 and 0.7 $M_2$ under axial flow entry situation or zero incidence. In Figure 5.2 and Figure 5.3 the surface Mach number distributions at different $Re_2$ are compared at 0.3 $M_2$ and 0.7 $M_2$ respectively under nominal incidence condition. In general, both the CFD and MISES seem to deliver very good agreement with the experiments at both extreme tested $M_2$ as well as different $Re_2$.

| No. | Parameter | $M_2$ | $Re_2$ x $10^6$ | Description |
|---|---|---|---|---|
| 1 | SMD | 0.3, 0.5, 0.7 | 1.0 | Inc. 0° |
| 2 | SMD | 0.3 | 0.4, 1.5 | Inc. 0° |
| 3 | SMD | 0.7 | 1.0, 1.5 | Inc. 0° |
| 4 | SMD | 0.3 | 0.4 | Inc. 0° |
| 5 | SMD | 0.7 | 1.0 | Inc. 0° |

Table 5.1 Nominal incidence conditions for test and simulation

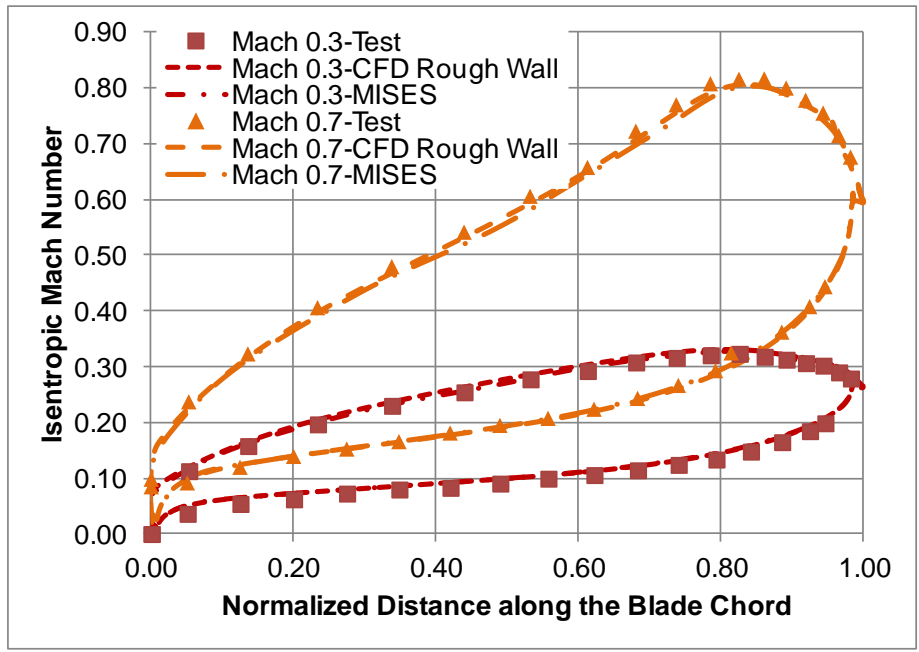

Figure 5.1 Isentropic surface Mach number distribution for $Re_2$ $1.0 \times 10^6$ at nominal incidence (0°) & $M_2$ 0.3 and 0.7

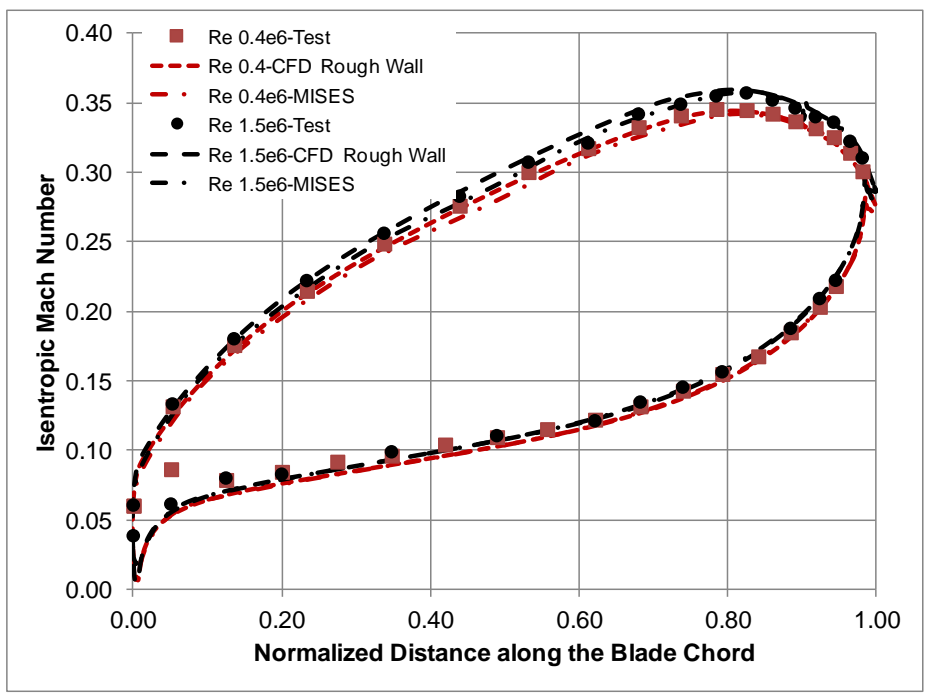

Figure 5.2 Isentropic surface Mach number distribution for different $Re_2$ at nominal incidence (0°) & $M_2$ 0.3

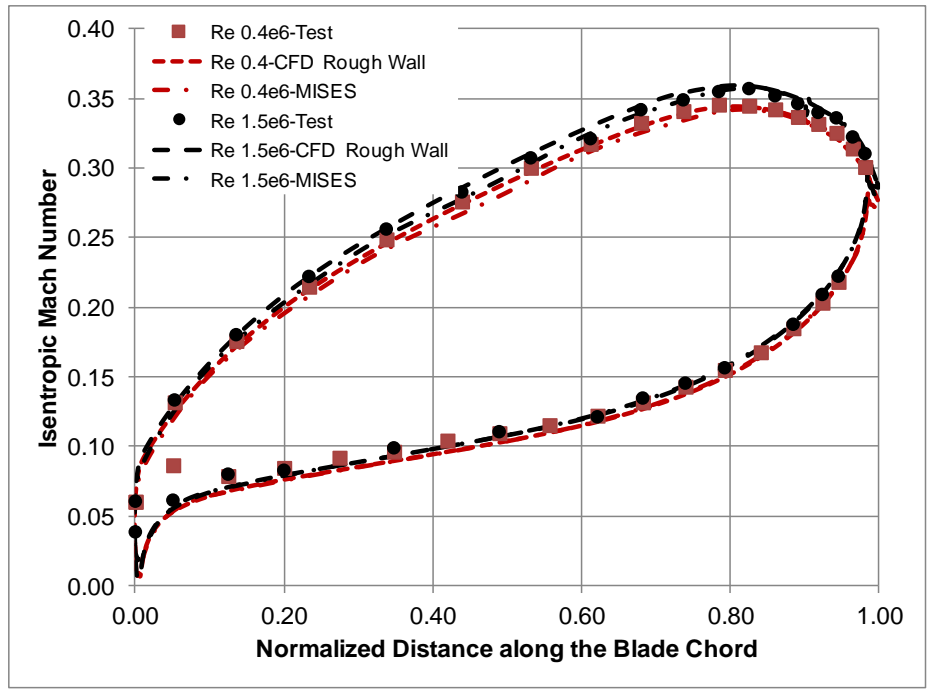

Figure 5.3 Isentropic surface Mach number distribution for different $Re_2$ at nominal incidence

$(0°)$ & $M_2$ 0.7

## 5.2. Analysis of section loading distribution at off-design incidence

| N° | Parameter | $M_2$ | $Re_2 \times 10^6$ | Description |
|---|---|---|---|---|
| 1 | SMD | 0.3 | 0.4 | Inc. -40°, +40° |
| 2 | SMD | 0.3 | 0.4 | Inc. -20° +20° |

Table 5.2 Off-nominal incidence conditions for test and simulation

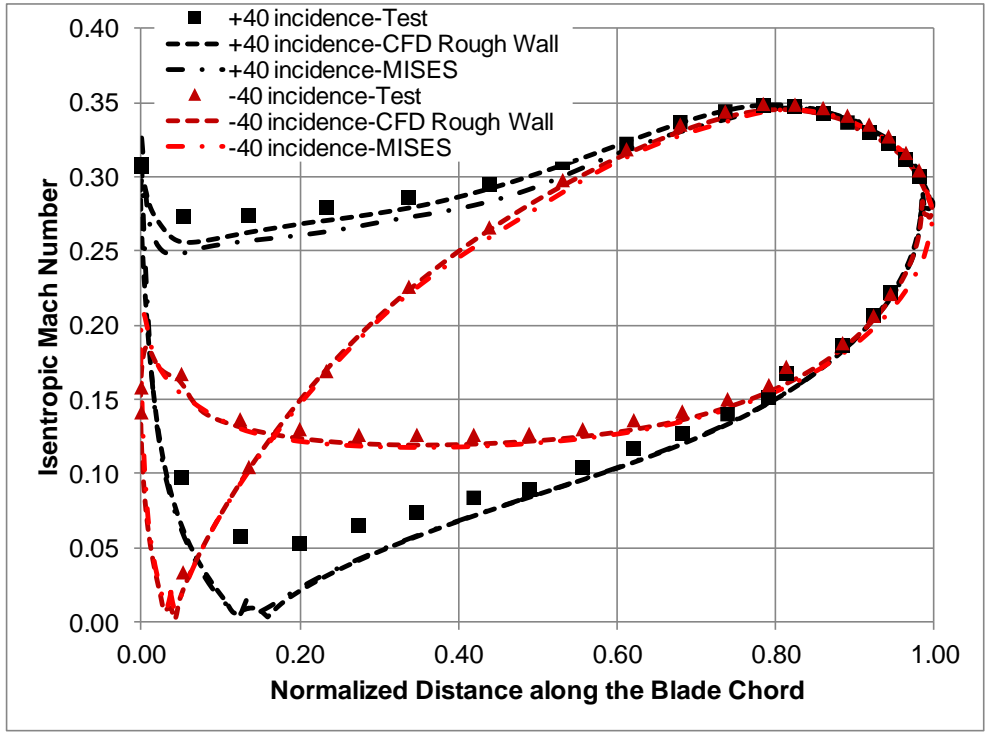

Figure 5.4 Isentropic surface Mach number distribution for $Re_2$ $0.4 \times 10^6$ at -40° & +40° incidence & $M_2$ 0.3

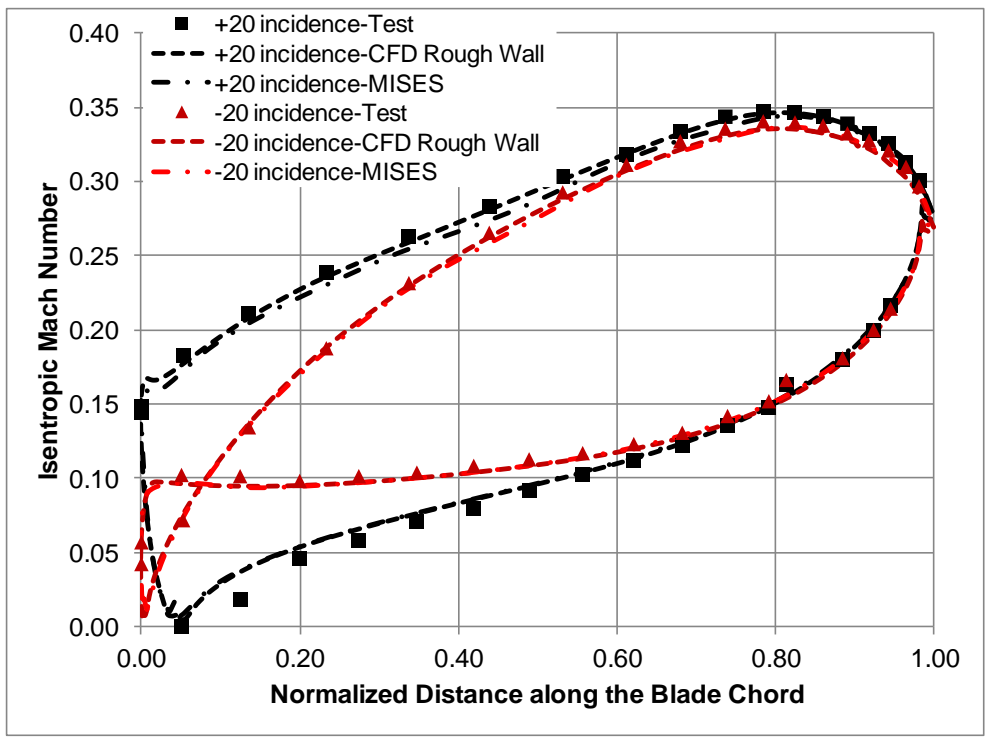

Figure 5.5 Isentropic surface Mach number distribution for $Re_2$ $0.4 \times 10^6$ at -20° & +20° incidence, $M_2$ 0.3

Predictions under large incidence conditions (± 40°) seem to result a slightly lower level of agreement (Figure 5.4 and Figure 5.5) but the involved discrepancies are within the measurement uncertainties.

*5.3. Analysis of section loading distribution with varying p/c and stagger angle*

| No. | Parameter | $M_2$ | $Re_2 \times 10^6$ | Description |
|---|---|---|---|---|
| 1 | SMD | 0.3 | 1.0 | *p/c*: 0.65, 0.82 &1.0 |
| 2 | SMD | 0.3 | 1.0 | $\theta$: nominal and extreme |

Table 5.3 p/c and stagger ($\theta$) conditions for test and simulation

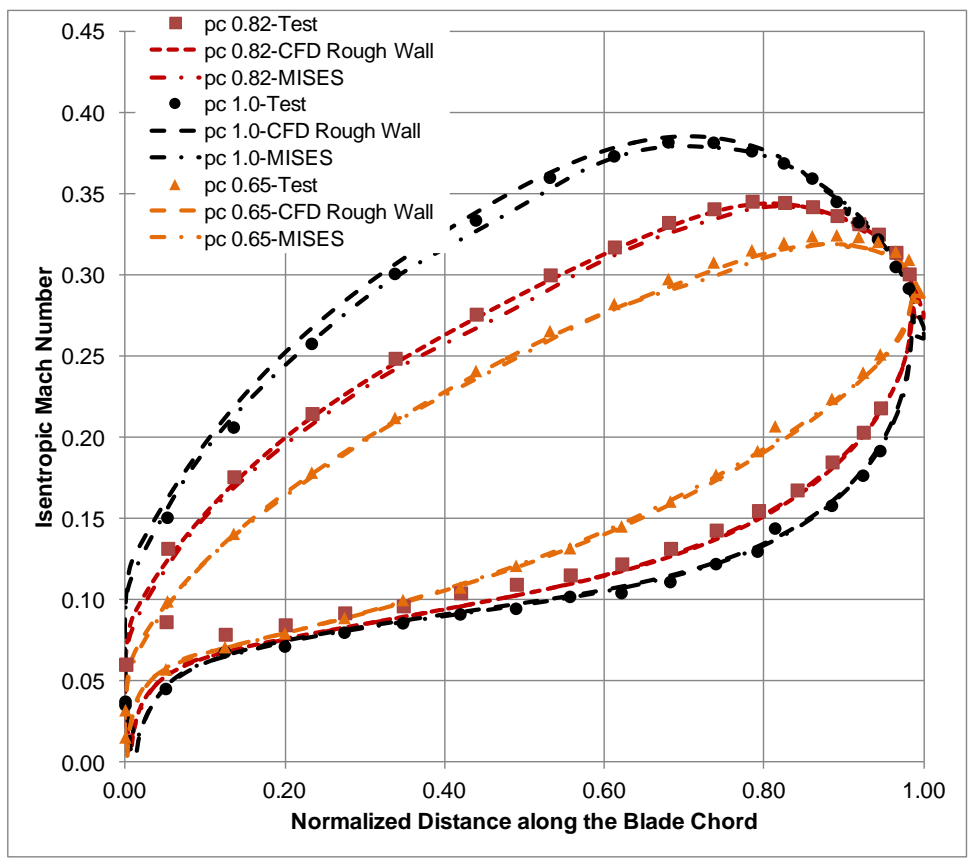

Figure 5.6 Isentropic surface Mach number distribution for $Re_2$ 0.4x$10^6$ at nominal incidence (0°) & $M_2$ 0.3 for p/c 0.65, 0.82 and 1.0

The target for *p/c* variation was to test the behavior of this profile section under largely differing loading characteristics and close to flow separation to which this profile section may be exposed. The additional *p/c* chosen were 0.65 and 1.0. For practical reasons, the stagger angle was kept unchanged, thereby the current variation of *p/c* entails a shift in gauge angle. Figure

5.6 shows the sectional loading associated with the *p/c* at nominal value of 0.82 together with 0.65 and 1.0. With increase in *p/c* the blade loading increases. The profile is well behaved with no flow separation even at extreme loading conditions. The intent of stagger variation has been to explore the performance of this blade section under significantly more closed gauge angle. The blade loading at nominal and extreme stagger is shown in Figure 5.7. Though the blade exhibits greater aft loading, the peak Mach number is like nominal condition, thus displaying the robustness of the blade.

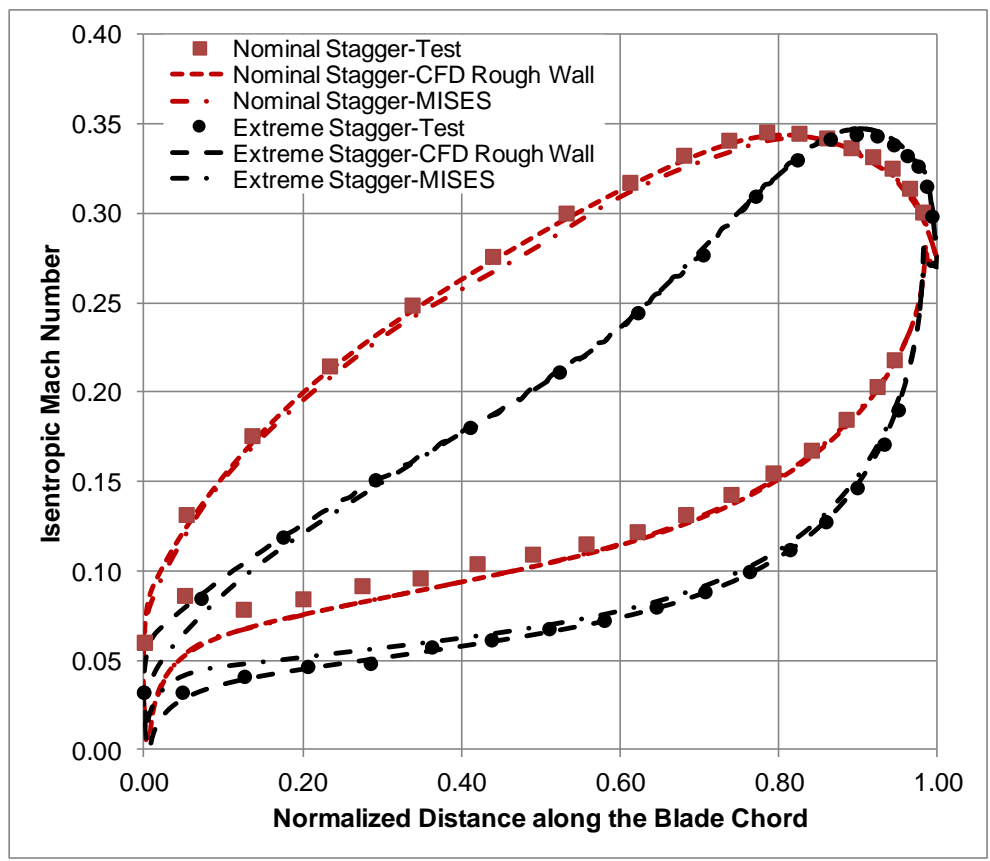

Figure 5.7 Isentropic surface Mach number distribution for $Re_2$ $1.0 \times 10^6$ at nominal incidence (0°) & $M_2$ 0.3 for nominal & closed stagger

*5.4. Analysis of profile loss & exit flow angle at nominal incidence*

| No. | Parameter | $M_2$ | $Re_2$ x $10^6$ | Description |
|---|---|---|---|---|
| 1 | $Y_2$ | 0.3 | 0.4 – 1.5 | ks=1.25, SMW, MISES |
| 2 | α | 0.3 | 0.4 – 1.5 | ks=1.25 , SMW, MISES |
| 3 | $Y_2$ | 0.7 | 0.4 – 2.5 | ks=1.25, SMW, MISES |
| 4 | α | 0.3 | 0.4 – 1.5 | ks=1.25 , SMW, MISES |

Table 5.4 Nominal conditions for test and simulation

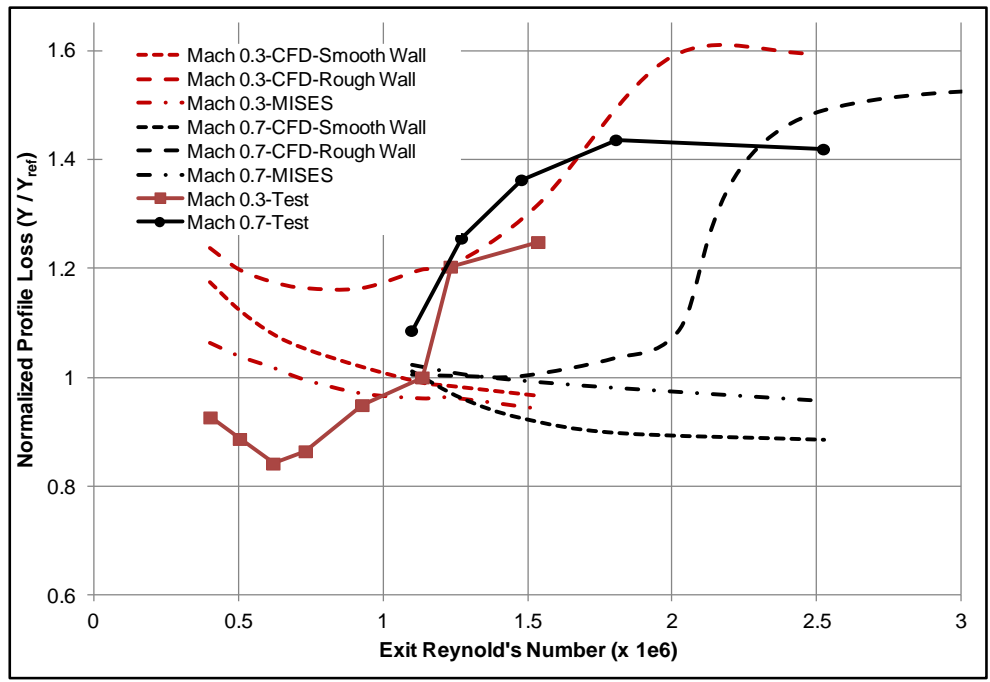

Figure 5.8 Nominal incidence normalized profile loss variation at 0.3 and 0.7 $M_2$

The normalized profile loss variation with exit Reynold's number for $M_2$ 0.3 and 0.7 is plotted in Figure 5.8. The experimental $Y_2$ is not very different for $M_2$ 0.3 and 0.7 and are withing the measurement error. The behavior of $Y_2$ can be explained if the variation with $Re_2$ can divided into three regimes: laminar, transition and turbulent. For 0.3 $M_2$, the laminar zone may be expected to span from 0.4 million to 0.6 million $Re_2$, transition zone from 0.6 million to 1.4 million $Re_2$, and fully turbulent beyond 1.4 million $Re_2$. In the laminar regime, $Y_2$ reduces with increasing $Re_2$ owing to reducing boundary layer thickness or shape factor at the blade trailing edge as seen in Figure 5.9. For 0.3 $M_2$, H reduces from 1.63 at 0.45 million $Re_2$ to 1.54 at 1.5 million $Re_2$. Even though laminar to turbulent transition occurs much earlier (0.2 mm from leading edge of the normalized blade chord at 0.45 million $Re_2$ to 0.05 mm at 1.5 million $Re_2$), the shape factor at the trailing edge becomes the determining factor for the profile loss. The profile loss in the fully turbulent regime asymptotes out as clearly observed at 0.7 $M_2$. The transition from laminar to fully turbulent regime in the test is primarily influenced by blade surface roughness.

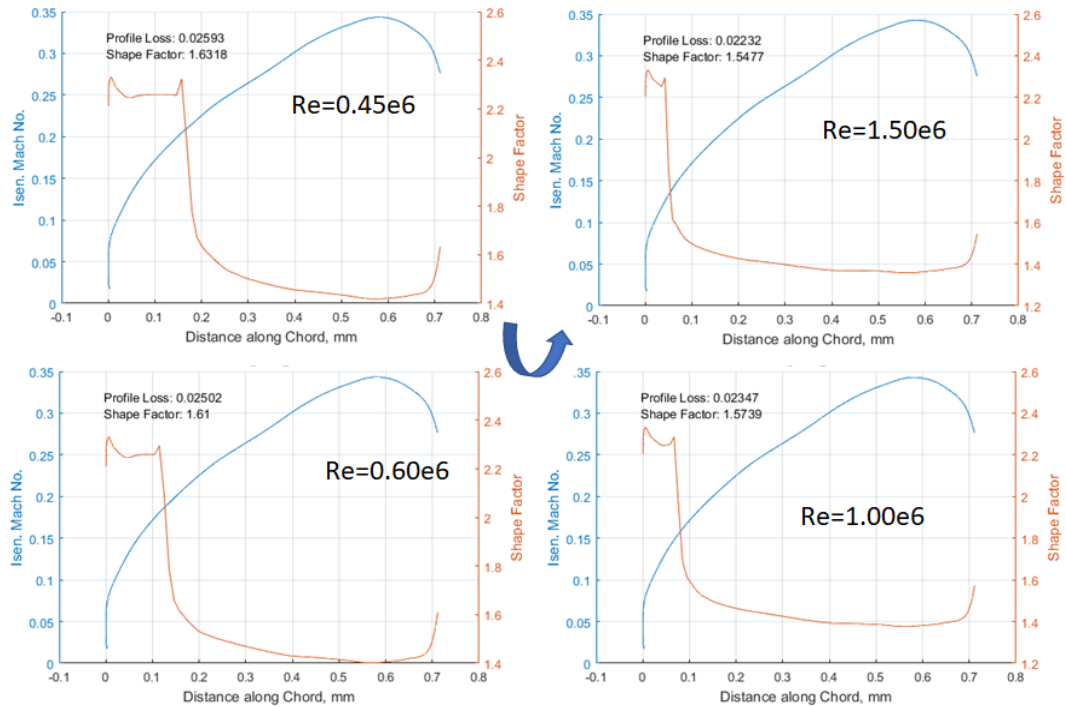

Figure 5.9 Shape factor variation with $Re_2$ at 0.3 $M_2$ for the suction surface

CFD result with roughness effect and transition seems to deliver reasonably good agreement at 0.3 $M_2$ under fully turbulent flow conditions but give rise to largest deviation in the low Re regime which is predominantly laminar. The smooth wall CFD and MISES results are comparable since MISES does not incorporate roughness correlations but does include transition model. At 0.7 $M_2$ and 2.5 million $Re_2$ the CFD result with rough wall matches closely to experiment, but smooth wall CFD and MISES deviate from the experimental result. The CFD with roughness and transition effect is the best model among the numerical schemes which closely approximates the test results but needs adaptation of its transition model to match with the tests in the transition regime. The flow deviation angle ($\alpha$) for 0.3 $M_2$ is plotted in Figure 5.10 and for 0.7 $M_2$ in Figure 5.11. Experiments show negative flow deviation

(additional flow turning in the unguided zone) and CFD indicates more negative deviation angle than measurements. CFD with roughness and transition shows closer agreement with the test results like the profile loss.

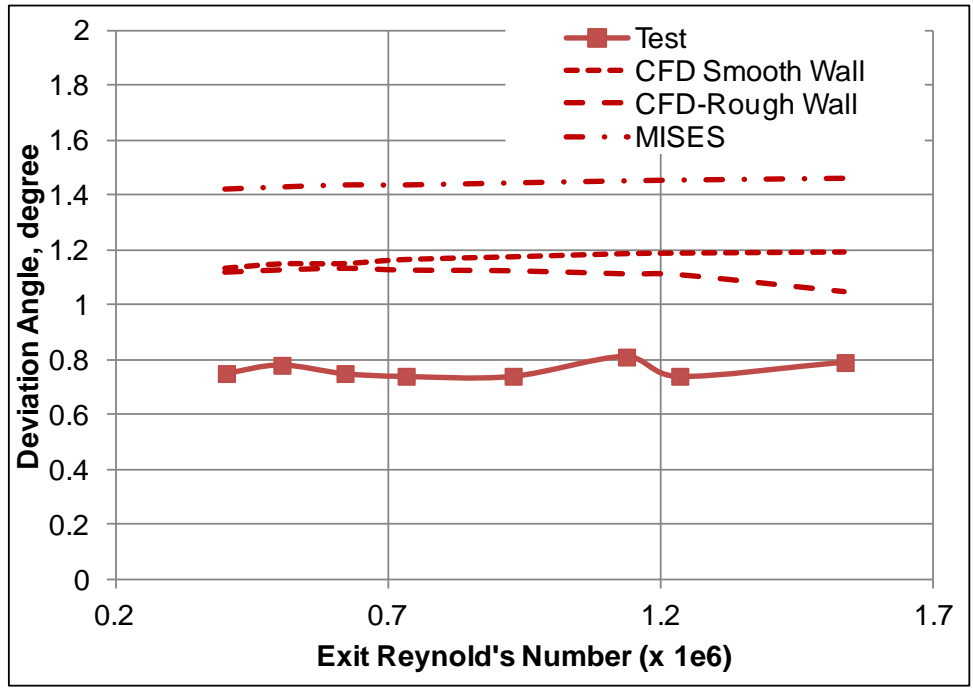

Figure 5.10 Deviation angle (α) variation with $Re_2$ at 0.3 $M_2$ and nominal incidence

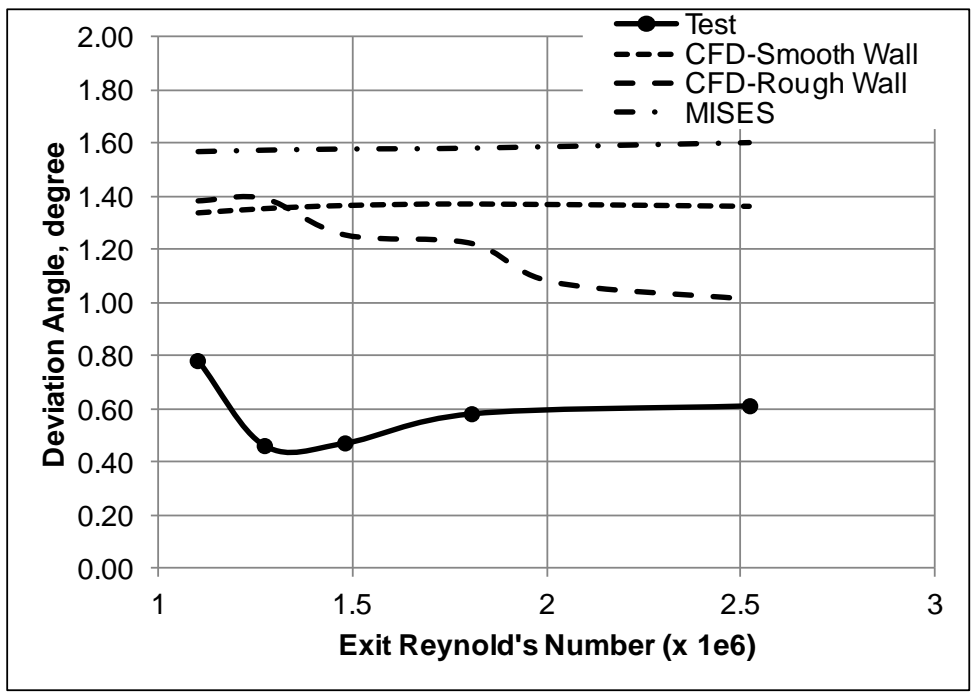

Figure 5.11 Deviation angle (α) variation with $Re_2$ at 0.7 $M_2$ and nominal incidence

## 5.5. Analysis of profile loss & exit flow angle at off-design incidence

The variation of profile loss and deviation angle for off-nominal conditions is presented at 0.3 $M_2$ and $Re_2$ varying between 0.4 million and 1.4 million as indicated in Table 5.5.

| No. | Parameter | $M_2$ | $Re_2 \times 10^6$ | Description |
|---|---|---|---|---|
| 1 | $Y_2$ | 0.3 | 0.4, 1.0, 1.4 | Inc. Variation |
| 2 | α | 0.3 | 0.4, 1.0, 1.4 | Inc. Variation |

Table 5.5 Off-nominal conditions for test and simulation

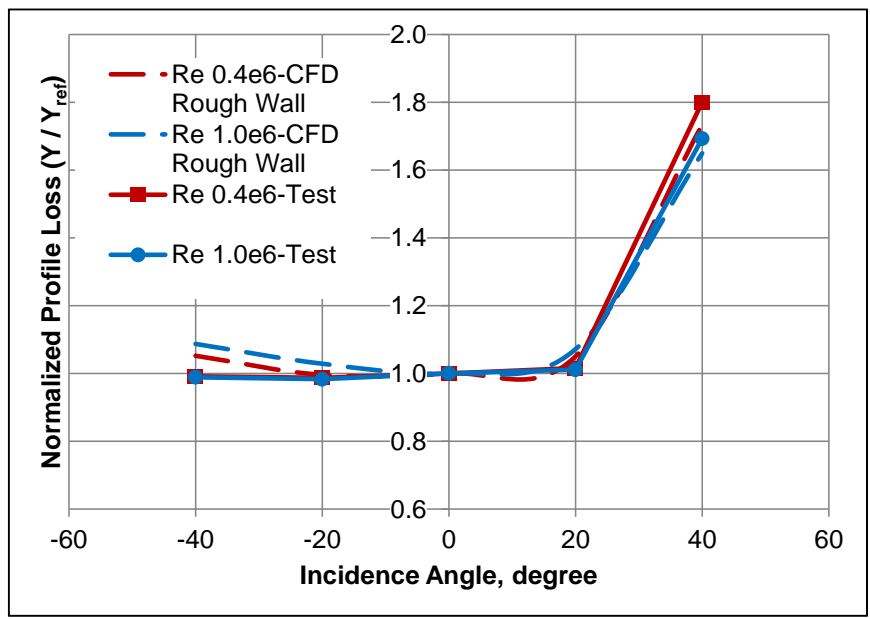

Figure 5.12 Off-nominal incidence normalized profile loss variation at 0.3 $M_2$

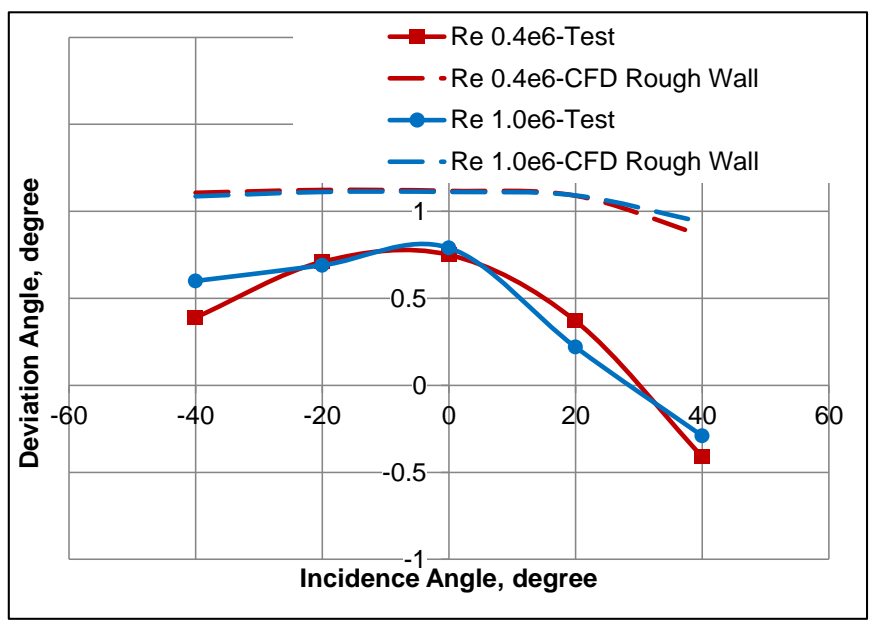

Figure 5.13 Off-nominal incidence deviation angle (α) variation at 0.3 $M_2$

Figure 5.12 and Figure 5.13 highlight normalized $Y_2$ and $\alpha$ variation with off-nominal incidence. $Y_2$ is lower for negative incidence and higher for positive incidence as expected. Experiment features a sizable variation in $\alpha$ for varying incidence angle. This can be attributed to a thickening of the boundary layer at the trailing edge which tends to diminish the flow turning in the unguided zone.

*5.6. Analysis of profile loss & exit flow angle at different p/c & stagger angle*

| No. | Parameter | $M_2$ | $Re_2$ x $10^6$ | Description |
|---|---|---|---|---|
| 1 | $Y_2$ | 0.3 | 0.4 – 1.7 | *p/c* variation |
| 2 | α | 0.3 | 0.4 – 1.7 | *p/c* variation |
| 3 | $Y_2$ | 0.3 | 0.4 – 1.7 | θ variation |
| 4 | α | 0.3 | 0.4 – 1.7 | θ variation |

Table 5.6 p/c and θ conditions for test

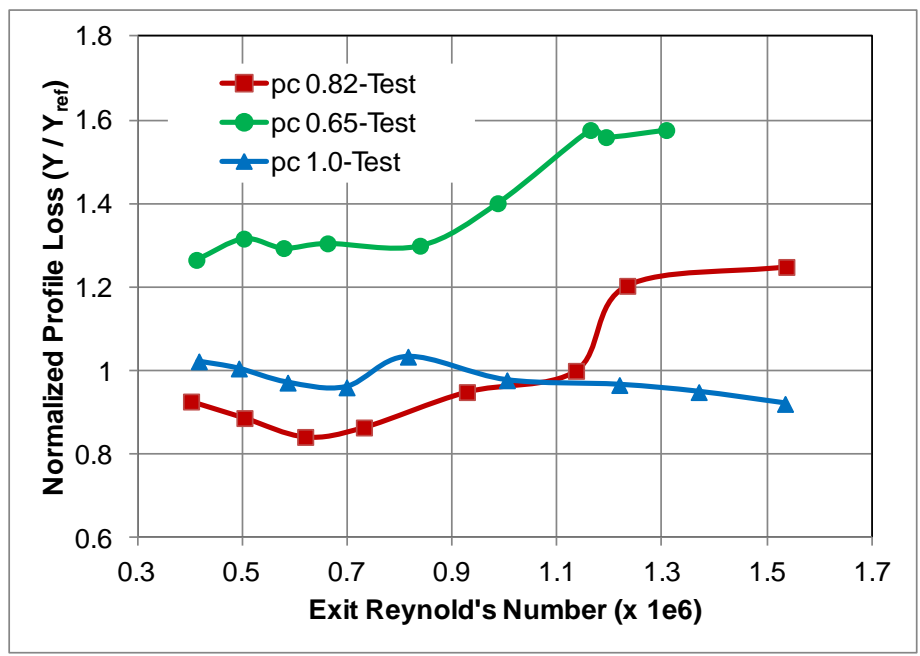

Figure 5.14 p/c influenced normalized profile loss variation at 0.3 $M_2$

The test conditions for *p/c* and stagger (*θ*) angle variation are shown in Table 5.6. There are two counteracting effects at higher *p/c*: reduced blade surface area due to lower blade count, variation of the gauge angle as the stagger angle is kept the same as nominal. The reduced blade surface area (wetted area) reduces blade friction and hence reduces profile loss but with a risk of flow separation near the blade tip. Since the stagger angle is not altered, the gauge angle slightly opens compared to the nominal gauge angle. From Figure 5.14, it is evident that the blade is well behaved as $Y_2$ values for higher *p/c* 1.0 is closer to the values at nominal *p/c* 0.82. On the contrary, the wetted area is higher for lower *p/c* and the blade becomes more closed. This results in lower profile loss for *p/c* 0.65 compared to the nominal *p/c*. However, the flow deviation angle ($α$) is significantly lower for p/c 0.65 compared to the nominal *p/c* as seen in Figure 5.15.

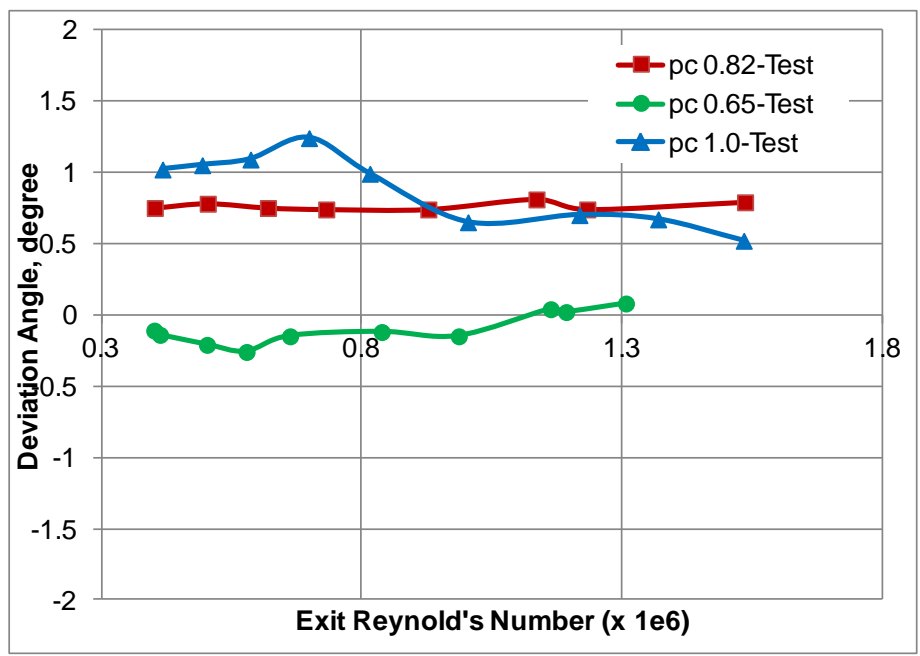

Figure 5.15 p/c influenced normalized deviation angle (α) variation at 0.3 $M_2$

The effect of extremely closed stagger angle (six degrees) on the profile loss is minimal as observed from Figure 5.16. This indicates the robustness of the profile to stagger angle variation. The deviation angle ($α$) is higher at extreme closed stagger angle compared to the nominal as shown in Figure 5.17.

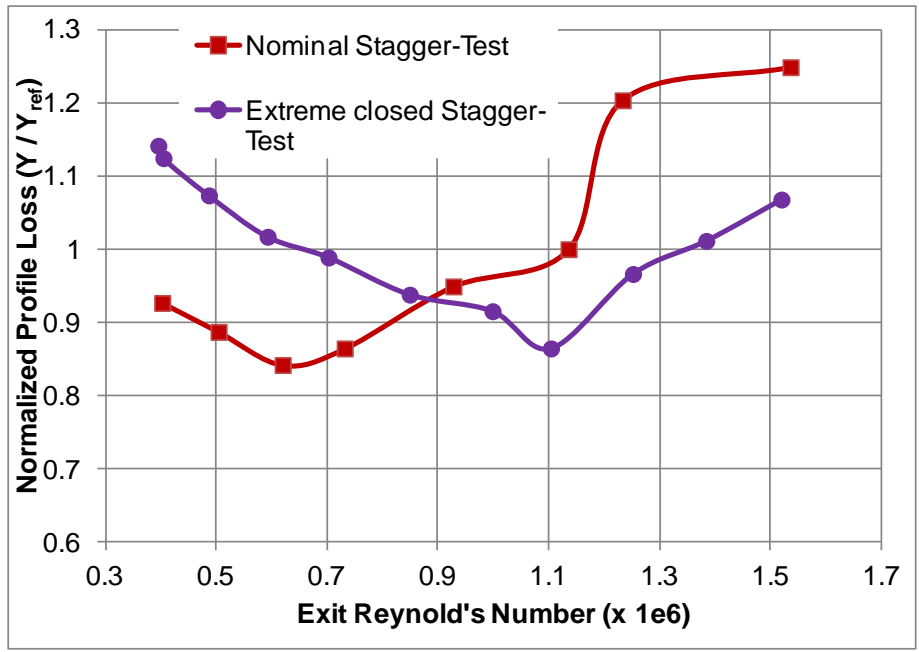

Figure 5.16 Stagger (θ) influenced normalized profile loss variation at 0.3 $M_2$

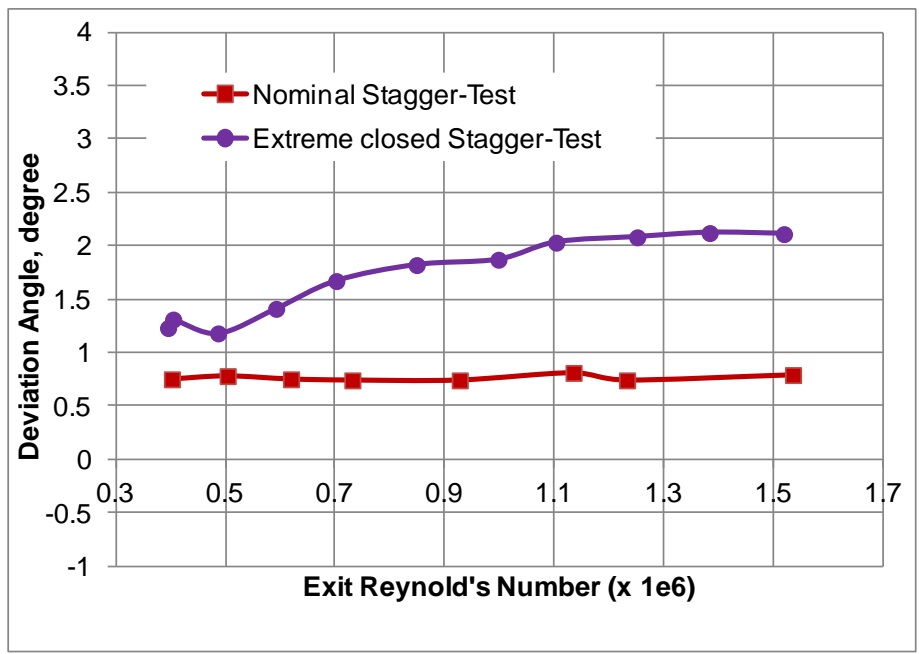

Figure 5.17 Stagger (θ) influenced normalized deviation angle (α) variation at 0.3 $M_2$

*Analysis of 3D loss and exit flow angle*

| No. | Parameter | $M_2$ | $Re_2 \times 10^6$ | Description |
|---|---|---|---|---|
| 1 | $Y_{2sec}$ | 0.3 | 0.4 | Span variation |
| 2 | $\alpha_{sec}$ | 0.3 | 0.4 | Span variation |

Table 5.7 3D test condition

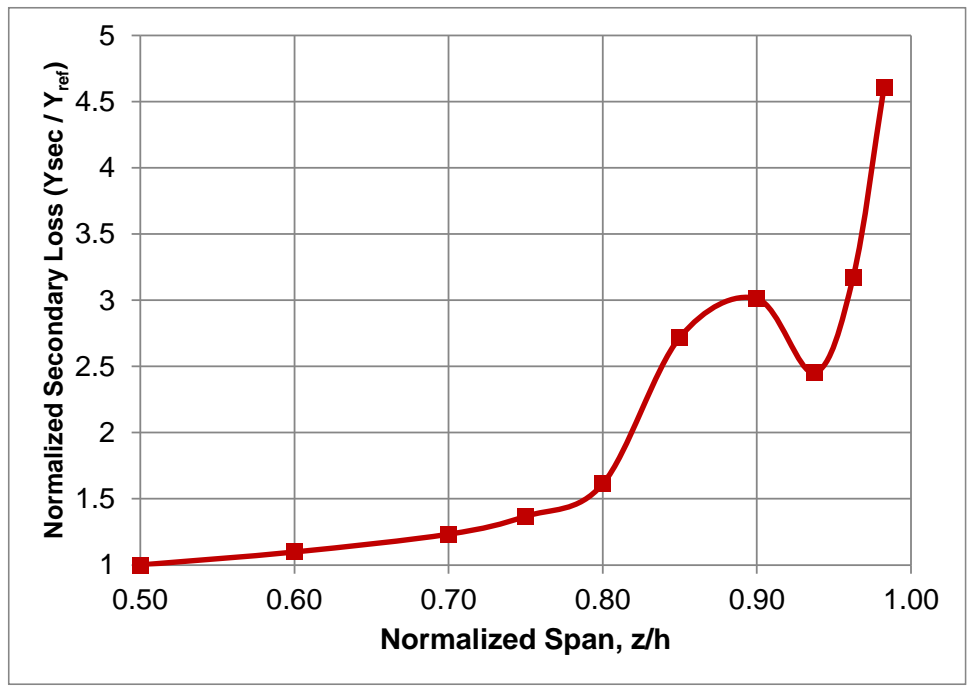

Figure 5.18 Span influenced normalized secondary loss variation at 0.3 $M_2$ *and 0.4*x$10^6$ $Re_2$

The 3D test at nominal incidence condition is carried out at 0.3 $M_2$ and 0.45 million $Re_2$ as shown in Table 5.7. The 3D test aims to assess the secondary loss which is the difference between the profile loss evaluated at a given spanwise location to that at the midspan. Thus, the secondary loss is null at profile midspan and only profile loss exists. The secondary deviation is the difference between outlet flow at a given location along the blade span to that at the midspan. Secondary deviation is zero at midspan like the secondary loss. The secondary loss increases exponentially from midspan to blade tip as shown in Figure 5.18. Secondary loss influence is highest near the blade tip due to flow vorticity (Figure 5.19 and Figure 5.20).

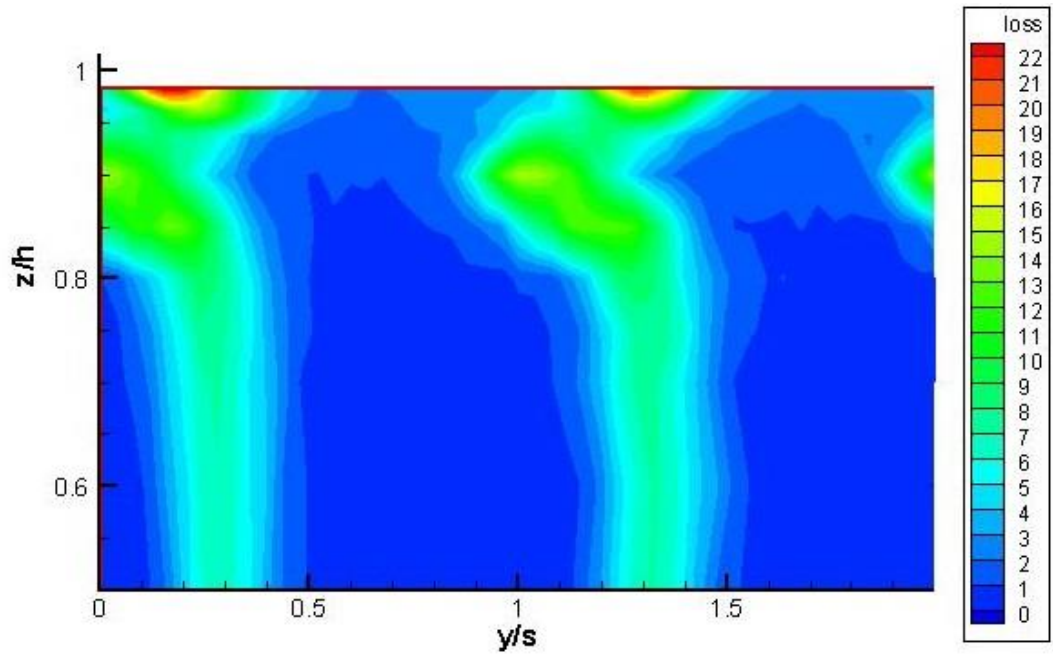

Figure 5.19 Total pressure loss from 3D test at 0.3 $M_2$ *and $0.4 \times 10^6$* $Re_2$

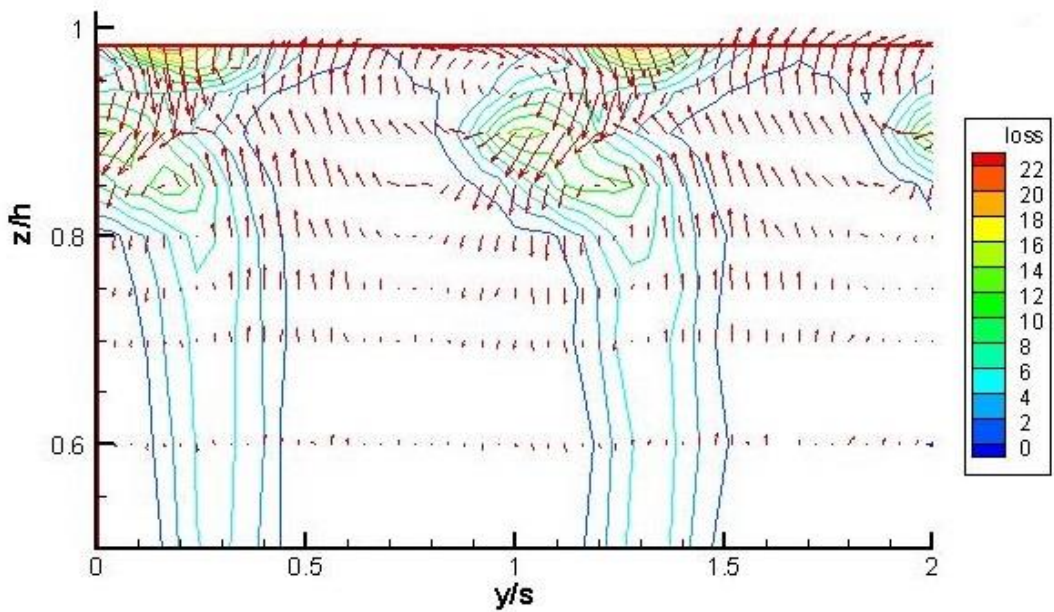

Figure 5.20 Total pressure loss & velocity vectors from 3D test at 0.3 $M_2$ *and $0.4 \times 10^6$* $Re_2$

The difference in the loss levels in bladed and un-bladed zone is clearly visible from the above plots. There is no flow separation anywhere along the blade span including the tip.

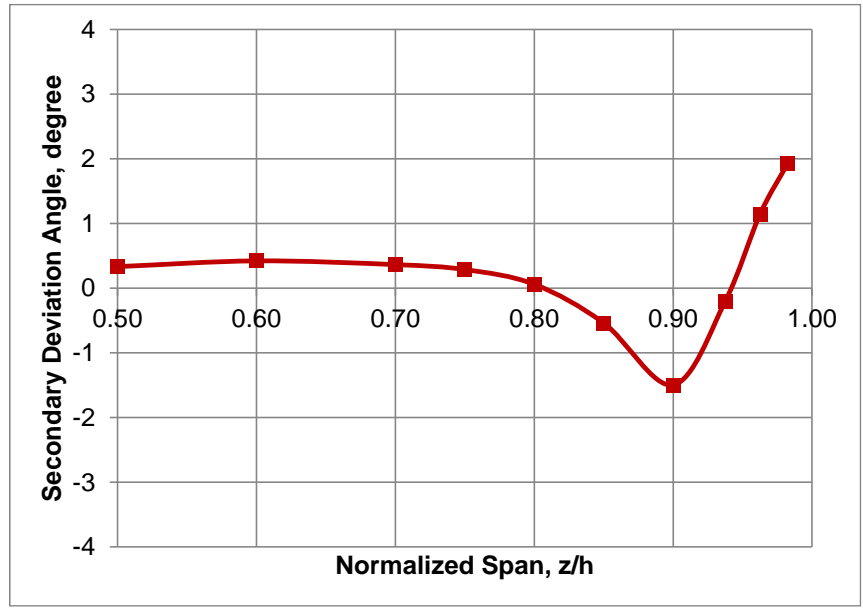

Figure 5.21 Span influenced secondary deviation angle variation at 0.3 $M_2$ and $0.4 \times 10^6$ $Re_2$

The secondary deviation angle starts to increase from 80% of the blade span and peaks at the blade tip as seen in Figure 5.21.

The 2D and 3D losses and deviation angles for experiment and numerical schemes at 0.3 $M_2$ and 0.4 million $Re_2$ is summarized in Table 5.8. The experimental 3D loss at nominal incidence is 3.08% indicating a good performance of the blade with combined profile and secondary losses. The experimental validation provides a sound basis for implementation of losses in the $sCO_2$ flowpath which is addressed in the next section.

| Parameter | Experiment | CFD-Rough Wall | MISES |
|---|---:|---:|---:|
| $Y_2$-2D | 1.64 | 2.10 | 1.80 |
| $Y_2$-3D | 3.08 | 4.42 | NA |
| α-2D, degree | 72.75 | 73.12 | 73.42 |
| α-3D, degree | 71.60 | 72.52 | NA |

Table 5.8 2D & 3D loss, deviation angle comparison at 0.3 $M_2$ & $0.4 \times 10^6$ $Re_2$

## 6. CONCLUSIONS

The experimental tests of the developed steam turbine blade was carried out at the Linear cascade Wind Tunnel facility at Politecnico di Milano, Italy. The test objectives were to estimate the blade profile loss, flow deviation angle, blade loading and 3D loss for an incidence range of -40° to +40°, Reynold's number variation between 0.5 million and 2.5 million, Mach number variation from 0.3 to 0.7, pitch-to-chord ratio variation, and stagger angle variation. The secondary objective of the test was to compare the test results with numeric solvers, MISES, and 3D CFD. The tests confirmed the increase in blade loading with Mach number. However, the blade loading does not change with Reynold's number. As the flow turning reduces with negative incidence the blade loading reduces, while owing to an increase in the flow turning, the blade loading increases for positive incidence. Increase in pitch-to-chord ratio increases blade loading as blade count per stage reduces to generate the lift equivalent to the nominal condition, and vice-versa for lower pitch-to-chord ratio. Even under extreme loading conditions the developed profile is well behaved with no flow separation. Both the numerical codes, MISES, and 3D CFD have shown no surface Mach distribution discrepancy at all the test conditions indicating the blade loading prediction accuracy of the solvers. The profile loss displays three regimes of operation namely laminar, transition and turbulent. In the laminar regime, profile loss has a minimum value of 1.5% at 0.3 $M_2$ and 0.6 million $Re_2$. It has a maximum value of 2.3% at 0.3 $M_2$ and 1.5 million $Re_2$, in the fully turbulent regime. In the laminar regime, the profile loss decreases with increasing Re due to, i) reduction in profile drag as the viscous sublayer of the boundary layer reduces, and ii) reduction in trailing wake mixing loss owing to shape factor reduction. The profile loss asymptotes in the fully turbulent regime as the profile drag loss reduction is negated by the increase in boundary layer loss due to turbulent eddies. In the transition regime the profile loss increases with increasing Re. This behavior of the profile loss in the transition regime is primarily due to blade surface roughness as it becomes the primary determining factor for boundary layer thickness. The MISES and smooth wall CFD show similar trends in profile loss as both do not account for surface roughness. The CFD with roughness and transition effect provides the best approximation to test as it shows similar trend and its value in the fully turbulent regime is comparable to test. However, the current level of CFD study is insufficient to explain the profile loss in the laminar and transition regime. The profile loss variation with incidence is like that reported from the design outcome. A key outcome from the cascade test is the invariance of the profile loss and deviation angle with Re at identical incidence angle. Lowest pitch-to-chord ratio exhibits higher loss due to increased wetted area. Cascade 3D loss is estimated by spanwise traversal of the 5-hole probe. The blade 3D loss is 3.08% at nominal incidence, 0.3 $M_2$ and 0.4 million $Re_2$. In summary, the linear cascade Wind Tunnel test validates the performance of the newly developed subsonic axial turbine blade for $sCO_2$ cycle application.

## References

[1] Song, B., Ng, W. F., Cotroneo, J. A., Hofer, D. C., and Siden, G., 2007, "Aerodynamic Design and Testing of Three Low Solidity Steam Turbine Nozzle Cascades," J. Turbomach., **129**(1), pp. 62–71.

[2] Jouini, D. B. M., Sjolander, S. A., and Moustapha, S. H., 2001, "Aerodynamic Performance of a Transonic Turbine Cascade at Off-Design Conditions," J. Turbomach., **123**(3), pp. 510–518.

[3] Mee, D. J., Baines, N. C., Oldfield, M. L. G., and Dickens, T. E., 1992, "An Examination of the Contributions to Loss on a Transonic Turbine Blade in Cascade," J. Turbomach., **114**(1), pp. 155–162.

[4] Hoheisel, H., Kiock, R., Lichtfuss, H. J., and Fottner, L., 1987, "Influence of Free-Stream Turbulence and Blade Pressure Gradient on Boundary Layer and Loss Behavior

of Turbine Cascades," J. Turbomach., **109**(2), pp. 210–219.

[5]  Bellucci, J., Rubechini, F., Arnone, A., Arcangeli, L., Maceli, N., and Dossena, V., 2012, "Optimization of a High-Pressure Steam Turbine Stage for a Wide Flow Coefficient Range," (44724), pp. 615–625.

[6]  Havakechian, S., and Greim, R., 1999, "Aerodynamic Design of 50% Reaction Steam Turbines," *Proceedings of the Institution of Mechanical Engineers, Part C: Journal of Mechanical Engineering Science*, pp. 1–25.

[7]  Youngren, H., and Drela, M., 1991, "Viscous/Inviscid Method for Preliminary Design of Transonic Cascades," 27th Jt. Propuls. Conf.

[8]  Persico, G., 2017, "Evolutionary Optimization of Centrifugal Nozzles for Organic Vapours," *Journal of Physics: Conference Series*, Institute of Physics Publishing.

[9]  Sathish, S., Kumar, P., Namburi, A, N., Swami, L., Fuetterer, C., and Gopi, P. C., 2019, "Novel Approaches for SCO2 Axial Turbine Design," Proc. ASME Turbo Expo 2019.

[10] Menter, F. R., 1994, "Two-Equation Eddy-Viscosity Turbulence Models for Engineering Applications," AIAA J., **32**(8), pp. 1598–1605.

[11] Langtry, R. B., and Menter, F. R., 2009, "Correlation-Based Transition Modeling for Unstructured Parallelized Computational Fluid Dynamics Codes," AIAA J., **47**(12), pp. 2894–2906.